\documentclass[11pt]{article}
\usepackage{graphicx}
 
\newcommand{\beq}{\begin{equation}}
\newcommand{\eeq}{\end{equation}}
\newcommand{\bea}{\begin{eqnarray}}
\newcommand{\eea}{\end{eqnarray}}

\begin{document}
\begin{center}
{\Large A realistic quasi-physical model of the 100 metre dash} \\
\vskip .25cm 
J.\ R.\ Mureika\thanks{newt@palmtree.physics.utoronto.ca} ~
\\
{\it Department of Physics} \\
{\it University of Toronto} \\
{\it Toronto, Ontario~~Canada~~M5S 1A7} \\
{\footnotesize PACS No.\ : Primary 01.80; Secondary: 02.60L}\\
\end{center}
{\small \noindent{\bf Keywords:} Mathematical modeling, sprinting, wind/altitude 
assistance, Track and Field}

\vskip .25 cm

\noindent
{\footnotesize
{\bf Abstract} \\
A quasi-physical  model (having both physical and mathematical roots)
of sprint performances is presented,
accounting for the influence of drag modification via wind and altitude
variations.  The race time corrections for both men and women sprinters are discussed, and 
theoretical estimates for the associated drag areas are presented.
The corrections are consistent with constant-wind estimates of previous authors, however
those for variable wind are more accentuated for this model.
As a practical example, the nullified World Record
and 1988 Olympic 100~m race of Ben Johnson is studied, and compared
with the present World Record of 9.79~s.  
}

\section{Introduction}

Mathematical models of athletic sprinting/running
performances, in some form or another,
date back to at least the early 1900s.  A.\ V.\ Hill \cite{hill}
was one of the pioneers to study a sprinter's ``velocity curve'', or
{\it i.e.} the runner's speed as a function of time.  
Mathematician Joseph Keller \cite{keller} formulated a simplistic model
to predict potential race times over a variety of distances.  His method
involved optimizing the set of coupled differential equations
\begin{eqnarray}
\dot{d}(t) & = & v(t) \nonumber \\
\dot{v}(t) & = &f(t) - \alpha\; v(t)~,
\label{tibshk}
\end{eqnarray}
with the race distance calculated as the integral
\begin{equation}
d = \int_0^T v(t)\, dt~.
\label{dist}
\end{equation}
In Equation~(\ref{tibshk}), $\alpha$ is a decay constant which 
places upper limitations on maximum velocities and velocity profiles.
Solutions are obtained by solving Equation~\ref{dist} subject to the 
initial conditions $v(0) = 0$, $d(0) = 0$.  Keller's original model
for sprint races proposed a constant propulsive force, $f(t) = F$, 
arguing 
that the athlete must use his/her entire strength to maximize their
performance.  In the works of Tibshirani \cite{tibs} and Mureika
\cite{mecjp}, the propulsive term is adjusted to vary with time, since
such intense physical exertion will inevitably introduce muscular fatigue.
Explicitly, a linear decrease was chosen, $f(t) = F - c\;t$, where
$c > 0$.  
 
\section{Quasi-physical model}

While Tibshirani's model yields an approximate match of final times,
it does not completely represent a simulation of an actual race.  A realistic
model of a sprint race should be able to accurately
reproduce the critical 10~m split data which is 
available\footnote{A ``split'' is defined as the elapsed time at each
10~m division of the race, or the time-interval in which a 10~m stretch
is covered.  The athlete's instantaneous velocities may be obtained for 
each 10~m increment, but due to technological and financial restrictions,
this type of data is currently rare.}.  In order
to better realize this approximation, the equation of motion
(\ref{tibshk}) is modified as follows:

\beq
\dot{v}(t) =  f_s + f_m - f_v - f_d~.
\label{mymodel}
\eeq
The term ``quasi-physical'' has been adopted to highlight the fact that
Equation~(\ref{mymodel}) is actually a mix of both mathematical and physical
components.  It is not a fully physical representation, nor is
is a purely mathematical model, although it can be used to effectively
study and estimate
key physical quantities (such as the elusive drag area of sprinters).  

The components of Equation~(\ref{mymodel}) 
are defined in the following subsections.

\subsection{Drive term $f_s$}
\label{fs}
A sprint race can be broken into roughly three different phases: the
drive, transition, and maintenance phases.  Previous models of sprint
performances ({\it e.g.} \cite{keller, tibs, mecjp} and references therein)
assign a singular propulsive term in the equation of motion.  These
do not explicitly account for the drive phase, in which a sprinter
begins a race from a crouched start.  From this position, the athlete
is able to achieve greater speed due to increased application of forces
(efficient drive posture).
This phase of the race lasts for only about the first 25-30 metres,
at which time the sprinter has transitioned to an upright running stance.

In the current model the following form of the drive term is 
adopted:

\beq
f_s = f_0 \exp(-\sigma\; t^2)~,
\label{driveterm}
\eeq
where $f_0$ is the magnitude of the drive, and $\sigma$ a constant to be
determined.  The $t^2$ dependence ensures that the term drops off rapidly
as the race progresses.  In fact, after roughly $t = 3$ seconds, the 
magnitude of the term drops to less than $0.1\%$ of its original value,
which would correspond roughly to the appropriate distance described
above.

\subsection{Maintenance term $f_m$}
This term is left relatively intact from the previous models, 
but assumes a slightly
different interpretation.  Once the drive term has dissipated, this term
is the only remaining propulsive component of Equation~\ref{mymodel}, and
represents the maintenance phase of the race.  
Explicitly,
\beq
f_m = f_1 \exp(- c\;t)~.
\label{maintenanceterm}
\eeq
Note that, unlike the linear maintenance term 
assumed in \cite{tibs, mecjp},
the time-dependence assumed herein is non-negative $\forall\;t$.
Certainly, this is a much more realistic assumption.

Due to the adoption of the drive term (\ref{driveterm}), the value of $f_1$
tends to be lower than predicted in previous models.  While this does not
immediately effect predictions for 100~metre performances, it does have
significant implications for 200~metre dash simulations (for which previous
models have made rather generous performance predictions; these are
discussed in detail in a forthcoming article \cite{me200}).  This difference
is further discussed in Section~\ref{propulsive}.

\subsection{Velocity term $f_v$}
\label{fv}
The first of two counter-propulsive terms, the velocity term is another
relic of the original Keller model, which accounts for a predicted 
$v(t)$ dependence in the equation of motion.  Such a term has some conceivable
physical interpretations: there must exist a physical barrier which limits
the maximum speed of a human, based more than just on pure muscular strength.  
It seems logical to assume that a sprinter's acceleration is 
curtailed with increasing speed ({\it e.g.} leg-turnover or stride rate
are physically and physiologically-constrained quantities).  The term is 
written
\beq
f_v = \alpha\; v(t)~,
\label{velocityterm}
\eeq
with $\alpha > 0$ a positive constant.  
A reasonable assumption might be that
the value of this parameter (based on the interpretation above) is
a ``physiological constant'' for most sprinters, and should 
not vary significantly from a prescribed value.  Most world-class sprinters
show stride-rates of about 4-5~strides/second (see {\it e.g.} the
analysis in \cite{seoul88}).

\subsection{Drag term $f_d$}
Along with the drive term, this is the most significant adjustment
to the model.  Explicitly,
\beq
f_d = \left(1-\frac{1}{4}\exp\{-\sigma\; t^2\}\right) \rho(H) A_d\;(v(t) - w)^2~.
\label{dragterm}
\eeq
The athlete is thus modeled as a thin slab of frontal cross-sectional area $A$, 
which is a sufficient approximation for the effects considered herein.
A critical factor in (\ref{dragterm}) is the ``modified drag area'' 
$A_d = C_d \cdot A/M$.  Here, 
$C_d$ the drag coefficient, and $M$ the sprinter's mass (since these models
are expressed in force per unit mass).  The exact values of these parameters
are unknown, since they can only be measured experimentally.  Previous
works \cite{davies,pritchard} have suggested that $A$ falls 
around $0.45$~m$^2$.
Additionally, these authors suggest that the drag coefficient
$C_d$ assumed a value on the order of $0.9-1.0$, but a recent suggestion
by Linthorne \cite{linth2} pegs the value as closer to $0.6$.  The
data obtained from this model is consistent with this statement, and in
fact indicates that $C_d$ may assume an even lower value
(see Section~\ref{results}).  

The expression $(1 - 1/4 \; \exp(-\sigma\; t^2))$ in (\ref{dragterm}) 
is designed as an initial correction to the cross-sectional area $A$,
representative of the transition from a crouched to upright position.  Although
the value $1/4$ is purely subjective, the actual magnitude of the correction
does not significantly affect the results.  It does, however, allow for 
a slightly greater acceleration over the first 10~m, and is a more
realistic assumption than having a constant $A$.

The important of an ambient wind in a sprint race is exemplified
by this term, since its magnitude is a function of $(v(t) - w)^2$.  Any
non-zero tail-wind reduces the effective drag experience by the sprinter,
allowing him/her greater acceleration, a greater top speed, and hence
a faster overall time.  A positive wind-speed corresponds to a tail-wind
({\it i.e.} wind in the direction of motion of the sprinter), while
a negative wind-speed denotes a head-wind.  The IAAF (International Amateur
Athletic Federation) has adopted a limit
of +2.0~ms$^{-1}$ for a supporting wind, above which a race is deemed 
``wind-assisted'' \cite{iaaf}.  Since such times are not recognized as
legal performances, they cannot be ratified as World Records.  

Wind-speed has the dominant effect on drag, but an additional factor which
affects this term is the atmospheric density, $\rho(H)$, where
\beq
\rho(H) = \rho_0 \exp(-0.000125\; H)~.
\eeq
Here, 
$\rho_0 = 1.184$~gm$^{-3}$ is the sea-level density at 25 degrees Celsius, and
$H$ is the measured elevation/altitude (in metres) (Dapena and Feltner
\cite{dapena} propose a second-order-in-$H$ correction to the altitude,
which is not included here).

Races which are run at altitudes above $H = 1000$~m are deemed 
altitude-assisted, but unlike wind-assisted marks, these can be ratified
as records.  For example, at the 1968 Olympics in Mexico City
($H \approx 2250$~m), World Records were set in both the 100~m and 200~m
dashes, thanks to the considerable altitude (the density of air in Mexico
City is roughly 76$\%$ of $\rho_0$).  Pietro Mennea's former 200~m record
of 19.72~s was also set at altitude.  Furthermore, the fastest-ever 
recorded 100~m clocking is 9.69~seconds by Obadele Thompson of Barbados,
in April 1996 (at which time the official World Record was 9.85~s).  
This race was run in El Paso, Texas ($H \approx 1300$~m) with a 
tail-wind of $+5.7$~ms$^{-1}$.  Thompson's previous 100~m performances 
were all slower than 10.00~s, which serves to demonstrate the 
extremal benefits of drag reduction.

\section{Model parameters and simulation results}
\label{results}
The coupled equations~(\ref{mymodel}) and $\dot{d}(t) = v(t)$ cannot be 
solved analytically, so one must resort to numerical methods.  Thanks to
the significant increase in modern processor speeds, the values of the
parameters $(f_0, \sigma)$; $(f_1, c)$; $\alpha$; $A_d$ can be rapidly 
isolated.  This was done using a fourth-fifth order Runge-Kutta 
integration scheme (written in C) run on a 500~MHz Pentium~III 
processor supporting  Linux~6.0.  An iterative time-step of 0.001~s was chosen
for the integration.

In order to determine extremal parameters for the model, key 10~m split
data for several world class 100~m races are matched.  Such data has been
obtained at various world-class track meets, including the 1997 and 1999
World Championships in Athletics \cite{athens97,sevilla99},
as well
as the 1988 Olympic Games \cite{seoul88}.  The instantaneous velocity splits
obtained in \cite{athens97} were measured with a laser-based device known
as a LAVEG, which collects its data by sensing the reflectivity of 
20~ns, 904~nm pulses directed at a (linearly) receding target.  
The cited reliability of the distance measurements is $\pm 1~$cm,
at a sampling rate of up to 100 measurements per second \cite{jenoptik}.

After repeated numerical runs and adjustments to the parameters,
the following set was obtained:
\begin{center}
$(f_0, \sigma) = (6.10, 2.22)$;\\
$(f_1, c, \alpha) = (5.15, 0.0385, 0.3225)$; \\
$A_d = 0.002875$;\\
\end{center}
Also, $w = 0.0$~ms$^{-1}$; $H = 0$~m.  The simulation results are
displayed in Table~\ref{modelsplits}, and yield a raw ({\it i.e.} excluding
reaction\footnote{Although they depend on the individual athlete and
the conditions at the time of the race, reaction times tend to range 
between $0.12 - 0.17$~s.  Reaction times below $0.100$~s are not allowed,
as it is believed that it is not physiologically possible to surpass this
limit.}) time of 9.70~s, with a maximum velocity of
$11.845$~ms$^{-1}$ at 59.18~m.  Such figures are in agreement with actual
world class sprinters running sub-9.90~second races to better than $\pm 0.01~s$,
with the instantaneous velocity splits also matching to within 1~$\%$ of the
actual value (see {\it e.g.} the data in \cite{athens97}).

Of course, each individual sprinter would most likely be described by a
unique set of parameters, and such a task is not within the immediate scope
of this paper.  Also, due to the nature of the drag term (\ref{dragterm}),
the effects of cross-winds are not considered in the current form of the model.

The drive-phase correction (\ref{driveterm}) is necessary to accurately 
reproduce
the observed velocity profile over the first 30-40~metres.  For example,
the velocity profiles presented in \cite{mecjp} for sprinter Donovan Bailey
of Canada give 9.32~ms$^{-1}$ (10~m), 10.95~ms$^{-1}$ (20~m), 11.67~ms$^{-1}$
(30~m), and 11.99~ms$^{-1}$ (40~m).  These velocity figures are doubtful,
in light of the profiles presented by actual split data \cite{athens97}.
It is unlikely that Bailey achieved a speed of
11.99~ms$^{-1}$ as early as 40~m, and furthermore sustained speeds in excess of
12~ms$^{-1}$ for an additional 40~m \cite{mecjp}.

\subsection{Wind assistance: determination of the drag coefficient and
frontal cross-sectional area}
\label{wind}
The value of $A_d$ is isolated in conjunction with the measured effect
of wind assistance by past authors \cite{keller,linth1,bb984},
most of 
whom agree that a tail-wind of $+2.0$~ms$^{-1}$ will
boost a 10~s 100~m sprint by about $0.1$~s.  The results of an earlier
study \cite{dapena} cite corrections of $-0.07~$s for a $+2$~ms$^{-1}$ wind, 
although one of the authors has since produced updated
results which are more commensurate with the literature \cite{dapena2}.
Table~\ref{modelsplits}
also shows the effects of such a tail-wind on the predicted 100~m times,
and accordingly predicts a boost of $0.104s$.  Conversely, a head-wind
of equal magnitude ($w = -2.0$~ms$^{-1}$) will increase the time by
$0.130$~s to 9.830~s.  Clearly, the non-linear nature of the drag term
implies that equal but opposite wind speeds will not provide equal
boosts.  It is reasonable to assume that there could be mild variations
in these corrections, depending on variations in $A_d$, but a full study of
this effect is not the aim of this paper.

For a sprinter of mass 80~kg, the value $A_d = 0.002875$ indicates that 
the effective drag area is $A \cdot C_d = 0.23$~m$^2$.  Note that this
is slightly less than the 0.3~m$^2$ estimate of Linthorne \cite{linth94}.
For a cross-sectional
area between $0.40-0.50$ m$^2$, this suggests that the drag coefficient
is between 0.46 and 0.57.  Note that it may actually be erroneous to assume
that the drag coefficient is constant, since the could potentially be
frequent transitions to laminar flow occurring throughout the race.  Also,
the drag area may vary slightly from athlete to athlete.  

It should be noted that while these simulation times are recorded to 
0.001~seconds,
official race times are reported to only 0.01~s, and official wind gauge
readings to 0.1~ms$^{-1}$.  In actual fact, electronic/photo-finish timing
is recorded to 0.001~s, and the following 
``rounding-up'' algorithm is applied to the times: unless the third decimal
place is 0, the hundredth position is rounded up.  For example, Canadian
sprinters Donovan Bailey and Bruny Surin have both recorded 100~m times
of 9.84~s, but their pre-rounded times were actually 9.835~s and 9.833~s
respectively \cite{bb984}.  A time of 9.831~s would be reported as 9.84~s,
while 9.830~s would earn a 9.83~s rounding.   

The reported value is measured from a gauge of height no more than
1.22~m, placed within 2~m from lane 1 on the in-field at a distance of
50~m from the finish line.  The wind-speed is sampled for a duration of
10~s from the start of the race \cite{iaaf}.  Wind speeds are measured 
to 0.01~ms$^{-1}$, but are rounded in a similar fashion.  However, Linthorne \cite{linth3}
has recently noted that measuring the wind speed in such a fashion
gives an error of $\pm$~0.7~ms$^{-1}$ on the actual value.  Such a discrepancy
would certainly impact the validity of the reported times.

\subsection{Altitude effects}
\label{altitude}
Although altitude effects are secondary insofar as drag modification
is concerned, their influence is not negligible.  Table~\ref{altsplits}
demonstrates the modifications to the times of Table~\ref{modelsplits},
subject to an increasing elevation.  Non-zero wind effects are not included,
but will be discussed in detail later on.  The model predicts an advantage
of $0.069$~seconds from sea level to a 2000~metre elevation.  For 
a ``Mexico City'' altitude (approximately 2250~m), the correction to this
100~m time (with no wind) 
would be $0.075$~s, in rough agreement with the predictions of
Linthorne \cite{linth1} and Dapena \cite{dapena2}.
It should be noted that the correction
figures given in Reference~\cite{bb984} are considered to be overestimated, due to
an inaccurate value of the drag area.  

\subsection{Graphical analysis of propulsive forces}
\label{propulsive}
While hypothesized from a mathematical basis, the quasi-physical model herein
provides accurate matches to real data.  A study of the propulsive forces
reveals a striking agreement to that presented by Dapena and Feltner
\cite{dapena}.  Although their analysis is for 10.90~s-caliber sprinters,
the figures presented herein for sub-10~s sprinters show the same overall 
structure.  The authors suggest that a plot of mass-normalized
propulsive force versus the athlete's velocity can be modeled by
two straight lines, adjoined at a ``boundary velocity'' roughly 3-5~ms$^{-1}$
less than maximum. 

\begin{figure} \begin{center} \leavevmode
\includegraphics[width=1.0\textwidth]{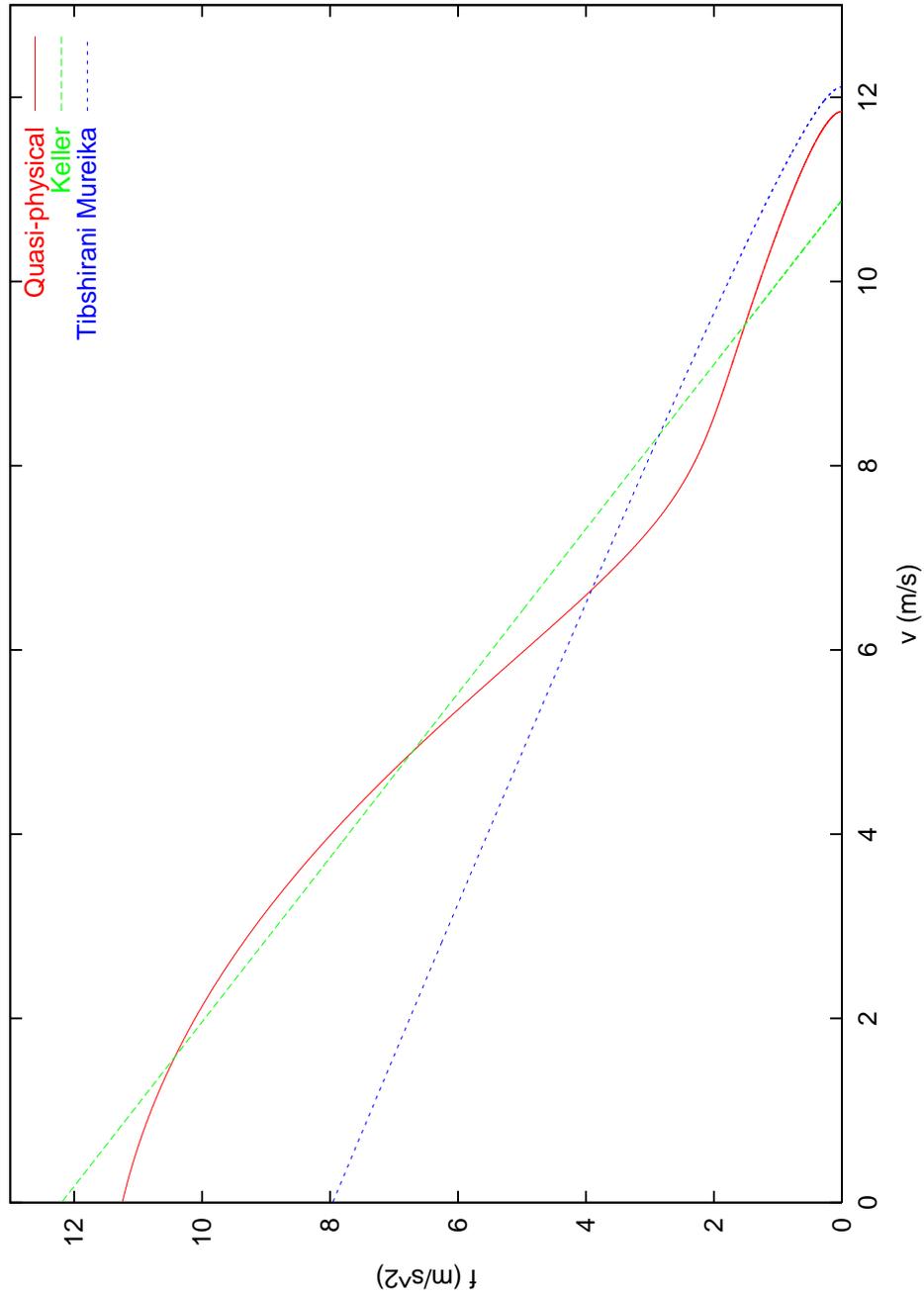}
\end{center} \caption{
Net positive propulsive force as a function of sprinter velocity.
Shown for comparison are the hypothetical World Record models of 
Keller (1973) and Mureika (1997) for 
Canadian athlete Donovan Bailey.  
}
\label{accel}
\end{figure}

Figure~\ref{accel} demonstrates the identical analysis for the model 
parameters listed in Section~\ref{results}.  Unlike Dapena and Feltner's
model, the acceleration is highly non-linear before 4~ms$^{-1}$, at which point
it assumes a roughly linear form to about 7~ms$^{-1}$.  The 
``boundary velocity''
in this case is representative of the point where the drive term 
$f_d$ has dropped to roughly 10~\% of its original value (about 1~second
into the race).  At this point, the
curve assumes another linear form of differing slope, before approaching
the maximum velocity (and again diverging from linearity).  

Such an analysis can in fact be used to rule out the Keller model and its
variants.  Although Keller's value of $F$ was in the 
$12.1$~ms$^{-2}$ range \cite{keller}, the overall form of the function 
does not provide an adequate global match.

Similarly, using Tibshirani's modification \cite{tibs}, the statistical
fit of race data from Donovan Bailey's 1996 Olympic 100~m race \cite{mecjp}
predicted an initial $f(0) = F$ value of 7.96 (Tibshirani's own analysis
predicted a lower value of 6.41, almost half that of Keller).  These
models are also presented in Figure~\ref{accel}, and are representative
of roughly a 9.8-10~second race.
On their own these two approximations do not support the
predictions of the quasi-physical model, however it is interesting to note
that Keller's model before, and the Tibshirani-Mureika model after their
intersection are a rough first-order approximation to the data.  In fact,
this is similar to linear model proposed in \cite{dapena}.

\subsection{Assistance for women sprinters}
\label{womencorrections}
Women sprinters are certainly lighter than an average 80~kg, yet their
cross-sectional area is correspondingly smaller.  Similarly, their stride
rates tend to be lower.  Thus, a first-pass approximation for women can
be made by a mild adjustment to the parameters of Equation~(\ref{mymodel}).
Following the interpretation of Section~\ref{fv}, a lower stride
rate can be taken to correspond to higher $\alpha$.  Similarly, adjustments
to the propulsive terms are reasonably made.

By setting $f_0 = 5.60; f_1 = 4.57; c = 0.037; \alpha = 0.32$, and
$A_d = 0.00275~$m$^2$ (which corresponds to a drag area of 
0.18~m$^2$
for a mass of 65~kg), one can reproduce splits which roughly replicate
those observed by world class female sprinter Marion Jones (see
\cite{athens97}), and match the predicted wind and altitude advantages
of \cite{dapena2} and \cite{linth1}.
For the cited drag area, a drag coefficient between
0.5-0.6 would require a cross-sectional area of 0.30-0.36~m$^2$, 
roughly 80\% that of men (a reasonable approximation).

Using these values gives a raw time of 10.653~s.  At sea-level ($H = 0$~m),
an advantage of $0.117$~s is predicted for a tail-wind 
of $+2.0$~ms$^{-1}$ (10.536~s), and $0.086$~s for $H = 2500~$m.  These figures are
again in close agreement with those of \cite{dapena2}
and Linthorne \cite{linth1}, who predicts
a correction of $0.12 \pm 0.02$~s. 
A wind of $-2.0$~ms$^{-1}$ at $H = 0$~m will increase the time to 10.802~s, a
difference of $0.149$~s.

\begin{figure} \begin{center} \leavevmode
\includegraphics[width=1.0\textwidth]{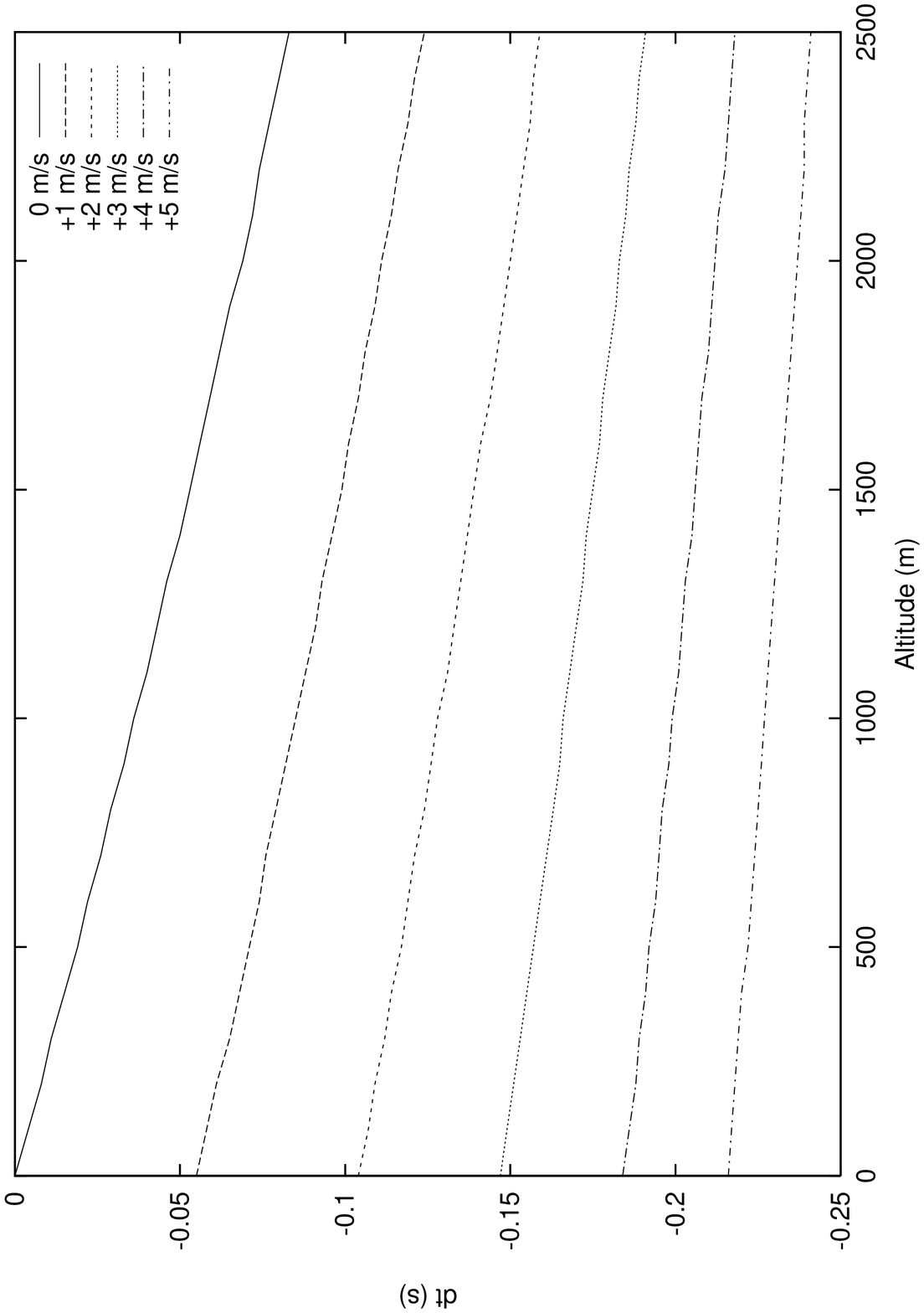}
\end{center} \caption{
World Class men's performance adjustments for ambient tail-winds
at varying altitudes.
}
\label{poswind}
\end{figure}

\begin{figure} \begin{center} \leavevmode
\includegraphics[width=1.0\textwidth]{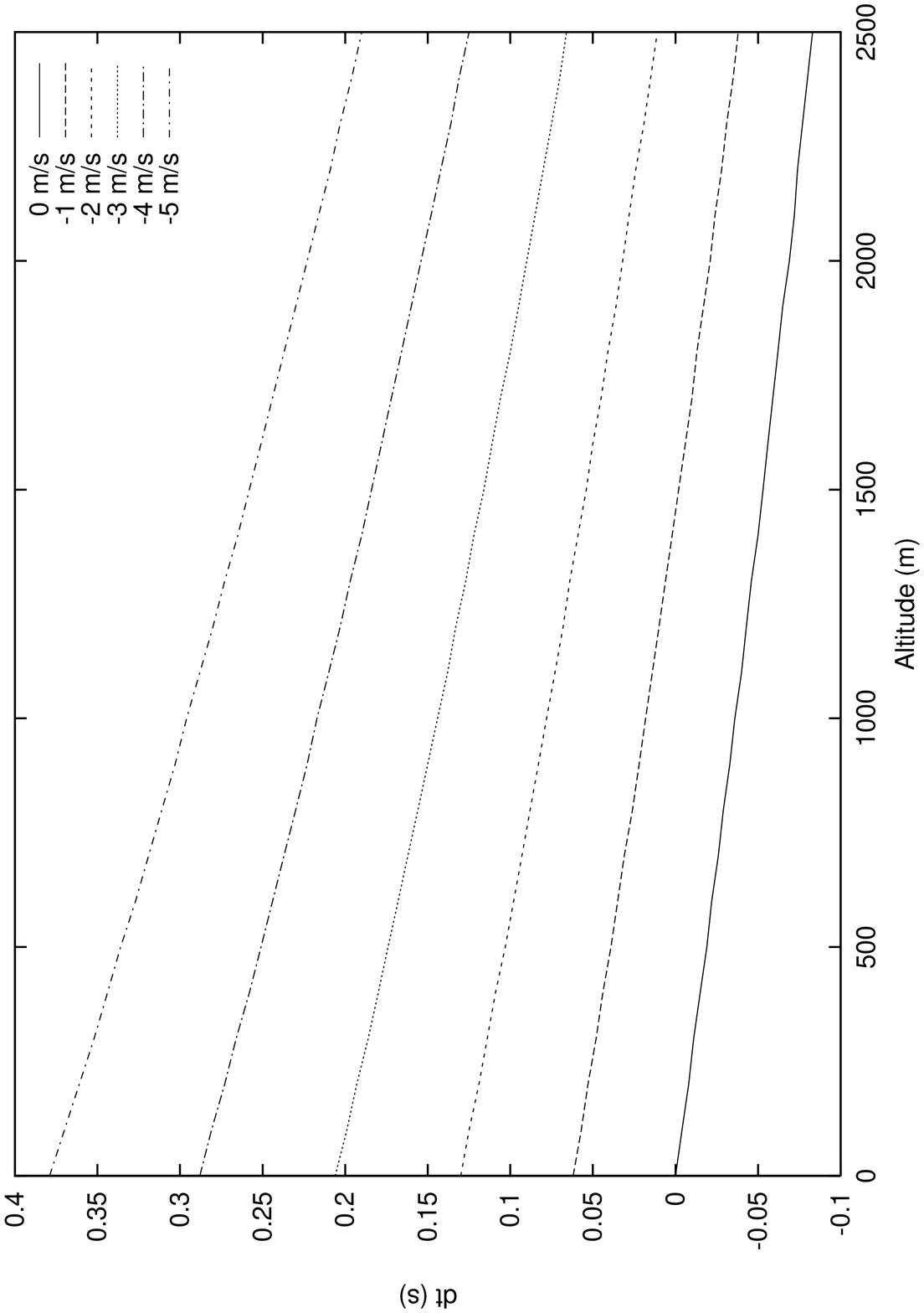}
\end{center} \caption{
World Class men's performance adjustments for ambient head-winds
at varying altitudes.
}
\label{negwind}
\end{figure}

\begin{figure} \begin{center} \leavevmode
\includegraphics[width=1.0\textwidth]{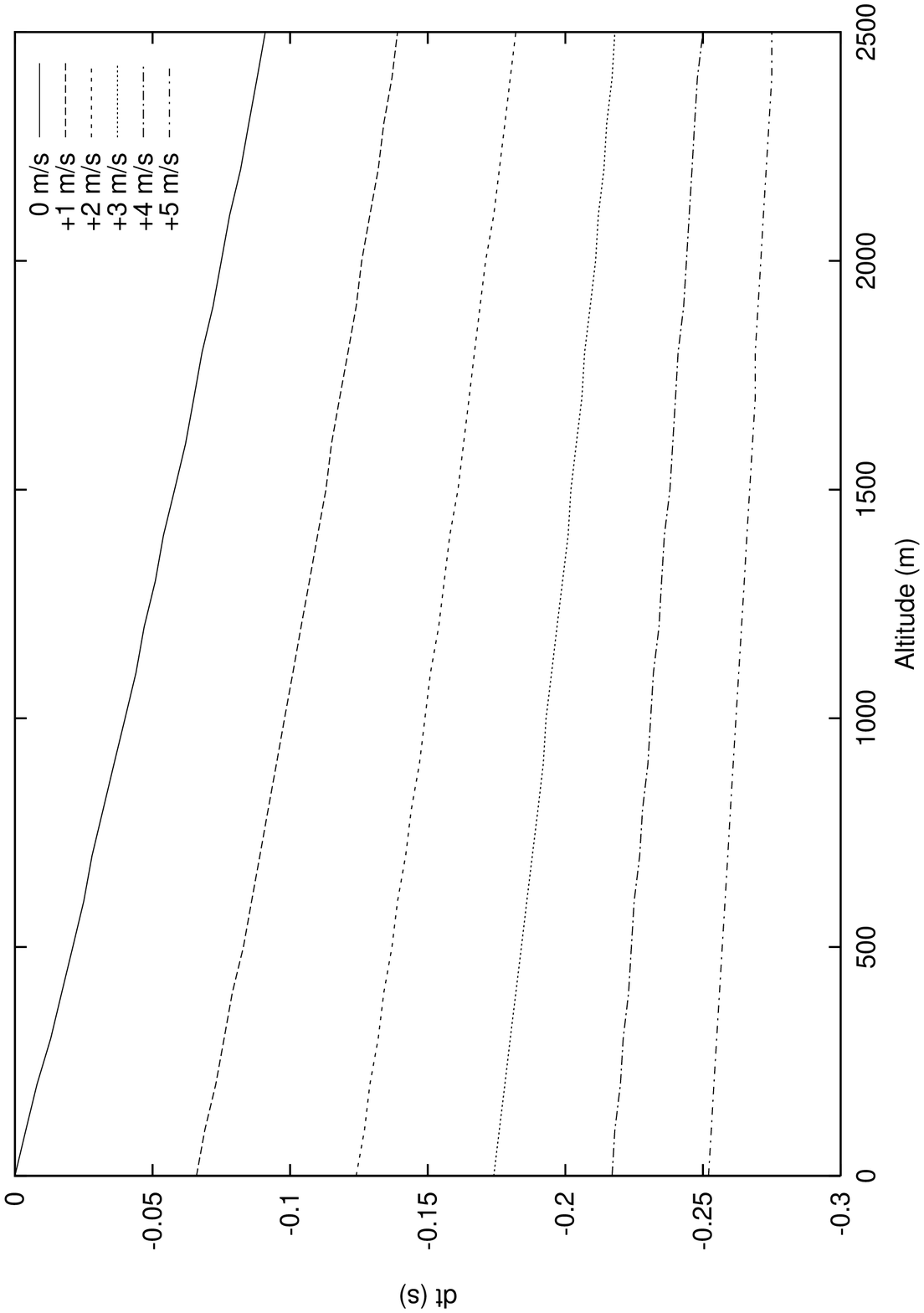}
\end{center} \caption{
World Class women's performance adjustments for ambient tail-winds
at varying altitudes.
}
\label{wposwind}
\end{figure}

\begin{figure} \begin{center} \leavevmode
\includegraphics[width=1.0\textwidth]{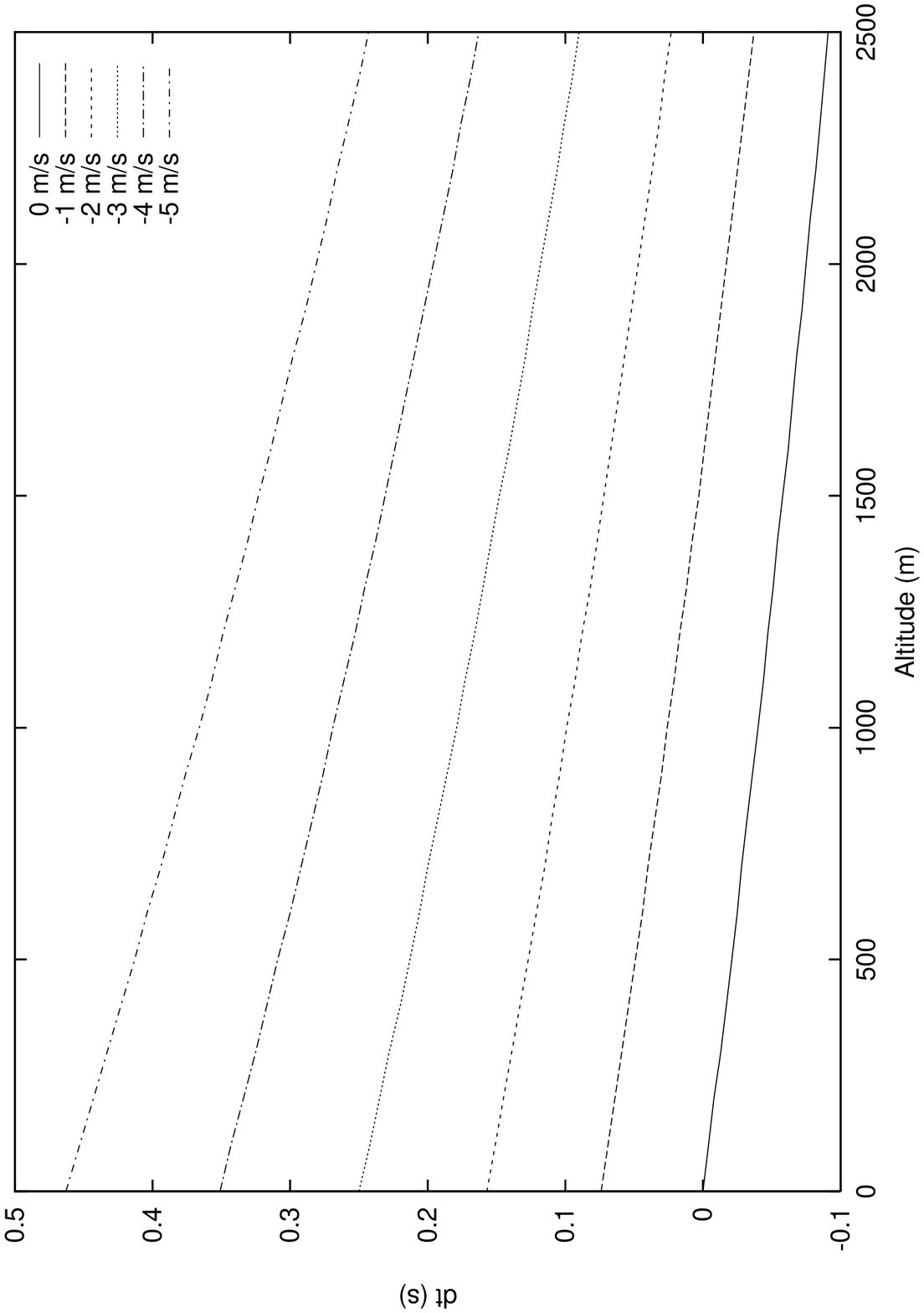}
\end{center} \caption{
World Class women's performance adjustments for ambient head-winds
at varying altitudes.
}
\label{wnegwind}
\end{figure}

Figures~\ref{poswind},\ref{negwind},\ref{wposwind}, and \ref{wnegwind}
give performance corrections for men and women,
for altitudes ranging between $H = 0 - 2500$~m.  The sign of the correction
indicates the impact of the associated wind and altitude on the {\it base}
time $t_0$ at 0~m altitude and 0-wind, {\it i.e.} $\Delta t = t_w - t_0$.  
A negative correction indicates
that the time $t_w$ is faster than the base ({\it e.g.} for tail winds), and
vice versa (note the change in sign relative to the Tables).   
The predicted corrections
for women cannot be applied to slower men's races, since the assumed drag
area is much lower.  If the modified drag area is kept as 0.002875~m$^2$kg$^{-1}$,
the resulting boost is $0.122$~s for a $+2.0~$ms$^{-1}$ wind and
a lapse of $0.154$~s for a $-2.0~$ms$^{-1}$ wind.

\section{Sensitivity to parameter variation}
\subsection{Correction for late-race velocity drop-off}
The measured velocity data presented in Reference~\cite{athens97}
indicate that the model predictions drop off faster than the measured
data for the latter half of the race.  It should be noted that this 
discrepancy can be accounted for by
lowering the power of $t$ in maintenance term (\ref{maintenanceterm}),
and hence the overall shape of the associated ``energy envelope''.
In fact, the author notes that a substitution $t \rightarrow t^\beta$,
with $\beta \sim 0.95-0.98$ (and mild adjustments to $f_1$) 
provides a closer match to the data.  

While such modifications are of
interest to providing a fully realistic simulation, variation of this 
parameter is an over-complication of an already-intricate model, and
is not essential for the study conducted herein.  A full study of such
time dependence is postponed for future work, and in particular its
implications for the 200~m dash are discussed in \cite{me200}.

Furthermore, it was noted that variation of the $(f_1, \alpha, c)$ 
parameters did
not significantly affect the velocity profile in the early part of the
race, but could be used to adjust the late-race profile.  Individual 
adjustments of these parameters to produce a 0.1~s
advantage in the final time presented minimal effects on the profile
for $d < 50$~m.  Conceivably, individual athletes could be represented
by their own set of parameters, which could help account for variability
in observed race times.  Despite these minor velocity adjustments over
60-100~m, a $+2.0$~ms$^{-1}$ assisting wind again provided boosts between $0.101-0.103$~s.

\subsection{Slow starts and the time dependence of $f_s$}
It is possible to use the model to simulate ``slow starts'' by suitable
readjustment of the drive term parameters.  In fact, by modifying the
time-dependence of Equation~\ref{driveterm}, one may reproduce starts
in which athletes did not accelerate as quickly as others, but achieved
higher overall velocities at each split in the earlier portions of the
race.

\begin{figure} \begin{center} \leavevmode
\includegraphics[width=1.0\textwidth]{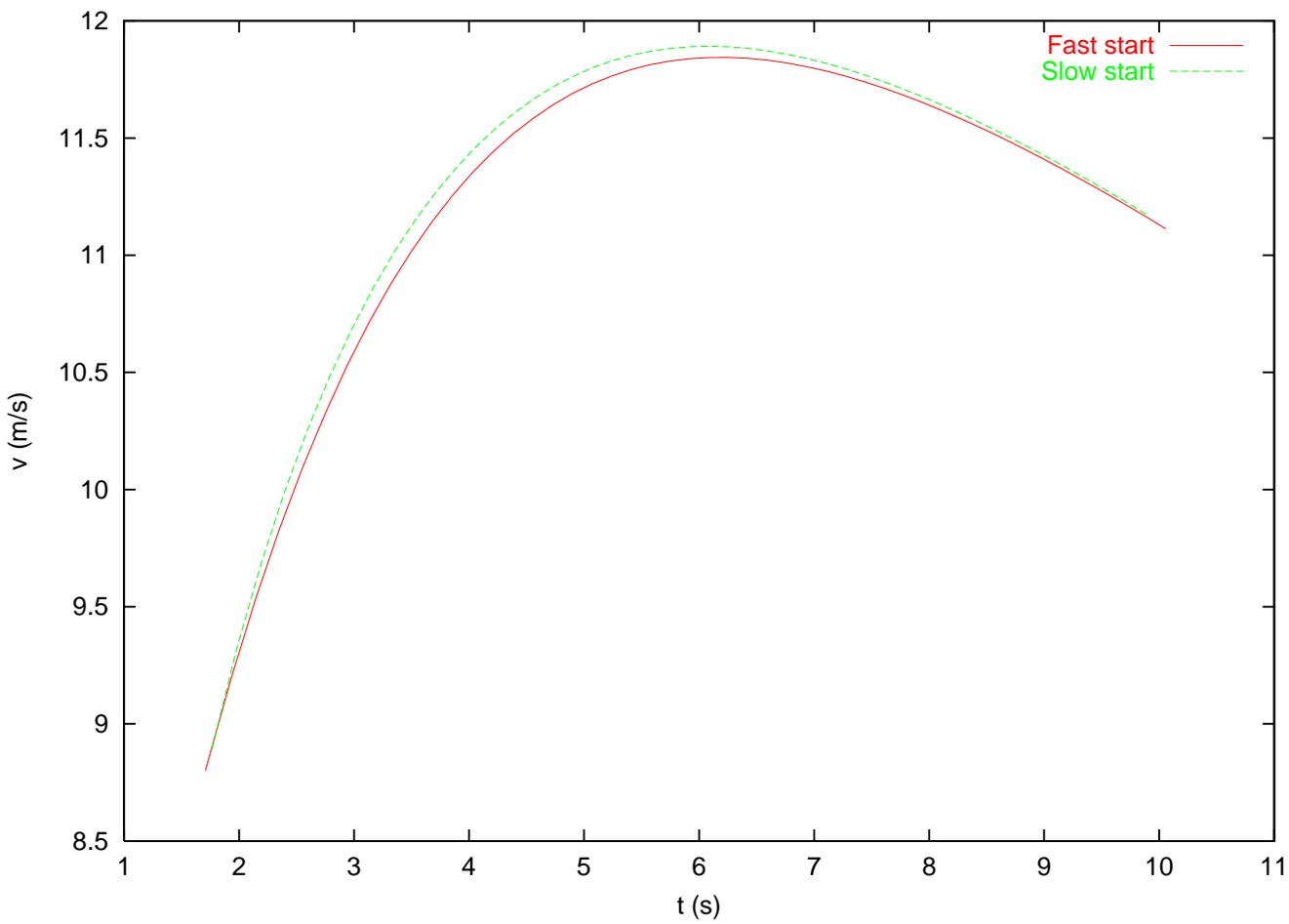}
\end{center} \caption{
Velocity-time profile for fast- and slow-start sprinter simulations.
}
\label{sf}
\end{figure}

Figure~\ref{sf} demonstrates such a simulation, with adjusted 
time-dependence $f_d = f_1 \exp(-\sigma \; t)$, and $\sigma = 1.68$,
as compared to the simulation of Section~\ref{results}.  The simulated
splits are shown in Table~\ref{bailey};
such data is consistent in profile with those
of Canadian sprinter Donovan Bailey from the 1997 World Championships
in Athens \cite{athens97}.
One can note this simulation lags behind the other
until about 70~metres, although achieving higher maximal velocities.
In this case, the finish is within 0.005~seconds.  This type of behavior
is frequently observable at any competition, and is characteristic of many
famous world class sprinters (including the aforementioned Bailey, as well
as American track legend Carl Lewis).

\section{Variable wind speeds}
\label{variablewinds}
As mentioned previously, the wind speeds are sampled for a period of 
10~seconds following the start of the race, after which the average value of the wind-speed is taken as
the official gauge reading.  The examples cited previously have assumed
that the wind velocity is constant.  However, this is probably more of
ideal situation than not.  A more realistic scenario is one in which the
wind speed is variable, but averages out to a value which may be 
unrepresentative of the true conditions.

Following the example of Dapena and Feltner \cite{dapena}, the results
of a variable-wind race are given in Table~\ref{varwinds}.  Four separate
variability conditions are simulated, with the constraint that the gauge
reading ({\it i.e.} the mean wind velocity over 10~seconds) be the
legal limit of $+2.0$~ms$^{-1}$.  These conditions include:

\begin{enumerate}
\item{Step function: $w(t) = 4 \;\Theta(t - 5)$}
\item{Step function: $w(t) = 4 [1 - \Theta(t - 5)]$}
\item{Linear: $w(t) = 2/5\; t$}
\item{Linear: $w(t) = 2/5\; (10 - t)$}
\end{enumerate}

Here, $\Theta(t)$ is the Heavyside function, with
$\Theta(t-t_0) = 0 \; \forall t < t_0$, and $\Theta(t-t_0) = 1$ otherwise.
The time-averaged wind-speed is calculated in the usual fashion,
${\bar w} =T^{-1}\; \int_0^T w(t)\;dt$, with $T = 10~$seconds.

While the constant tail-wind speed of $+2.0$~ms$^{-1}$ predicts a boost of
0.104~s for the present model, the above conditions predict a rather 
large range of variation.
Cases~1 and 3 show the smallest advantages, due to minimal
(or zero) wind conditions in the drive phase (hence lower overall 
accelerations in the first half of the race). 
These cases also
show the lowest peak velocities in this boost scenario, although still lower
than the constant wind case.  Note that the
peak velocities occur much later than in the base 
case {\it and} the constant-wind case.

Case~4 shows a peak velocity at roughly the same location as in the base
9.700~s run (a difference of only 19~cm), although with a much higher 
magnitude ($+0.182$~ms$^{-1}$).  Case~2, on the other hand, shows almost the
same velocity, but at a much earlier mark (54.95~m).
This decidedly premature maximum is no doubt due to the overpowering 
contributions of the velocity term $f_v$ once the assisting tail-wind 
has subsided.  The sprinter is physically unable to achieve a higher
velocity.

Dapena and Feltner only consider Cases~1 and 2 in their analysis (for
the time-averaged +2~ms$^{-1}$ wind).  However, it is interesting that the
predictions of this model in fact are opposite of their assertions.  For
Case~1, they predict a boost of 0.066~s, while their Case~2 shows a boost 
of 0.060~s, both lower than their constant-wind case (0.070~s).  The
conditions of Case~1 and Case~3 allow the maximum velocity to be achieved
later in the race, which may not have an issue addressed by the authors
in question.  Their underestimation of the initial boost may also be a
contributing factor, and adjustment of their parameters to suit the
current estimates of Dapena \cite{dapena2} may yield differing results.  Whether
this is a false estimate in the figures of Dapena and Feltner or a short-coming
of the current model is an open question.  Observation of this 
velocity-limitation could yield credibility to form of $f_v$.

Certainly, such wind speed variations considered in this section can further
contribute to small velocity discrepancies between the model with constant
wind speed, as well as variations in the actual race splits.  

\subsection{Variable 0-wind average}
The analysis above begs the question: what kind of performance variations 
could be expected for a time-averaged wind speed of 0~ms$^{-1}$?
Table~\ref{wind0avg} demonstrates such potential discrepancies, with linear
and sinusoidally-varying wind speed functions (the unrealistic step functions
are omitted here).  The linear winds simply range from equal-but-opposite
maximum/minimum values (with 0~ms$^{-1}$ occurring at the 5~s midpoint), the
cosine functions begin and end at the indicated maximum and minimum, and
the sine functions achieve a the maximum indicated with 0~ms$^{-1}$-endpoints.
Absolute maxima of 2 and 4~ms$^{-1}$ are chosen.

\begin{figure} \begin{center} \leavevmode
\includegraphics[width=1.0\textwidth]{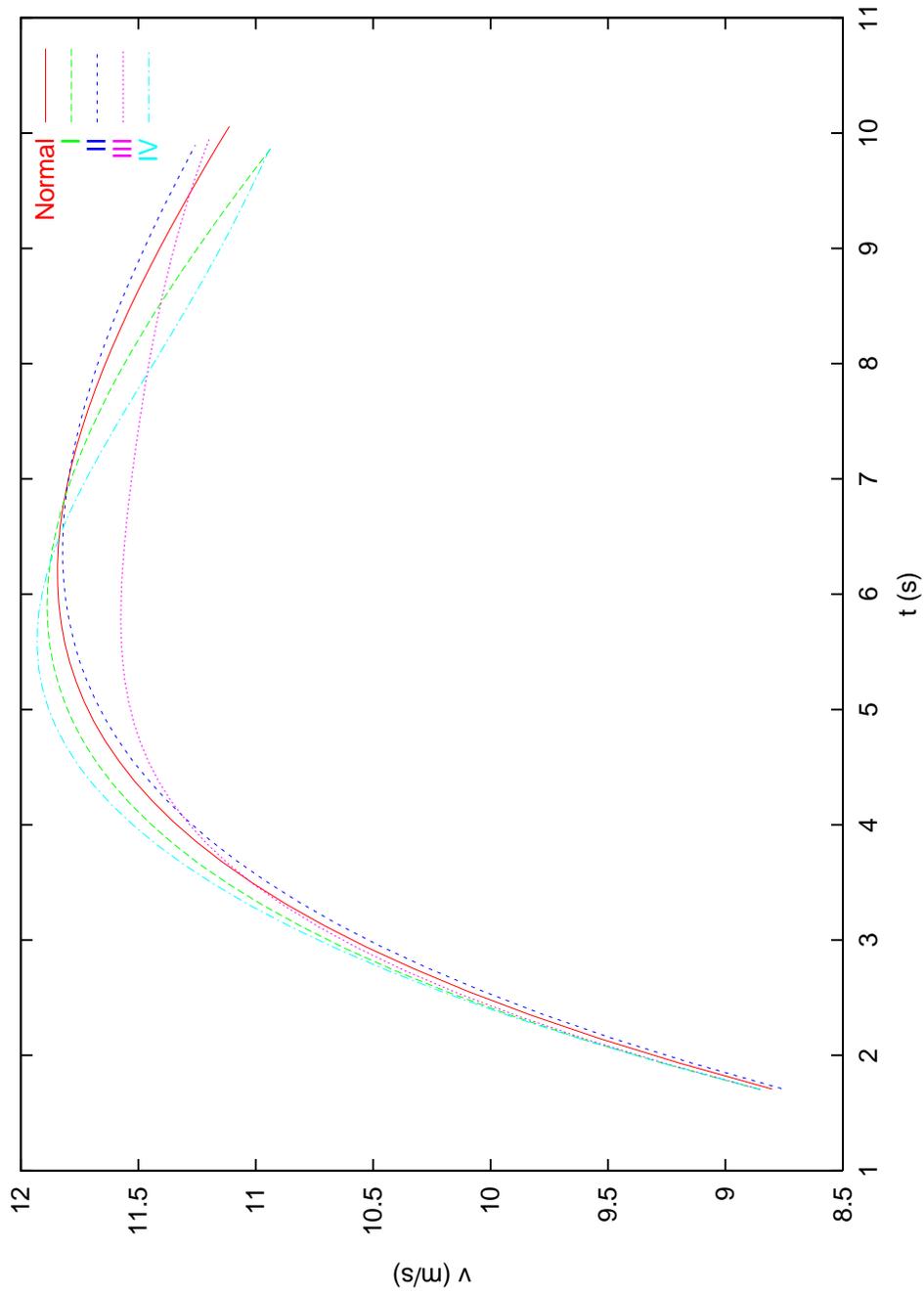}
\end{center} \caption{
Velocity profile for 0-average variable wind conditions.  Cases include:
I: $w = 4/5 \;(5-t)$; II: $w = 4/5 \;(t-5)$; III: $4\;\cos(\pi t/5)$;
IV: $4\;\sin(\pi t/5)$.
}
\label{vw}
\end{figure}

The simulations for the 2~ms$^{-1}$ absolute wind show variations between 0.034~s 
slower and +0.023~s faster than the actual 0~ms$^{-1}$ wind-speed values of 
Table~\ref{modelsplits}.  Interestingly enough, the greatest advantage comes
from a cosine wind which begins and ends as a head-wind, the form of which 
allows for greater acceleration through middle of the race.   Conversely,
the greatest disadvantage is from the equivalent negative-cosine wind.
As the overall extremum of the wind increases, the differences become more
acute (although note a much higher disadvantage when the wind is negative
from about 30~m onward).
Figure~\ref{vw} presents the velocity profiles of several of these variable 
conditions, compared to the base case of Section~\ref{results}.  

\section{Comparison to previous wind-correction estimates}
\label{comparison}
Recently, a ``back-of-the-envelope'' expression for 
potential wind and altitude corrections was obtained in \cite{bote}:

\beq
t_{0,0} \approx \left[1.027 - 0.027 \exp(-0.00125\;H) \;(1 - w \cdot t_{w,H} / 100)^2 \right] \; t_w~,
\label{wind2}
\eeq
Here, $t_{w,H}$ is the official race time 
(run with wind $w$ at altitude $H$), and $t_{0,0}$ the time for 
$w = 0$~ms$^{-1}$ at sea level.
This is derived in part from Equation~\ref{dragterm} by selecting a constant
propulsive force, and an average velocity $v = 100 / t_{w,H}$.  The assumption
is made that a sprinter expends roughly $\delta = 2.7\%$ of his/her 
energy fighting drag at sea level.  For non-zero $w, H$,
$\delta \sim \rho(H) A_d (v_{w,H} - w)^2$.
The effects of wind are thus calculated by assuming the adjusted drag
impacts the sprinter's velocity by the ratio of forces $F(v_{w,H}-w)/F(v_0)$, where 
$v_{0,0} \simeq 100/t_{0,0}$ is the average velocity at sea level with no wind
for a race-time $t_{0,0}$, with $v_{w,H} = 100/t_{w,H}$ the 
wind/altitude-influenced velocity/time.

This ready-to-use expression provides a good match to the predictions of 
this model, as well as
those of Dapena \cite{dapena2} and Linthorne \cite{linth1}.  
Table~\ref{boteTable} demonstrates the corrective potential of Equation~(\ref{wind2})
subject to the model parameters of Section~\ref{results}.  Note that the
approximation becomes slightly worse for increasing absolute wind speeds,
however the magnitude of these variations is less than 0.5\% for
$w \in [-5, +5]$~ms$^{-1}$.  Such errors can certainly be accounted for
by mis-estimations of the input variables.

The mismatch may also result from the constant velocity assumption in 
the derivation.  The interested reader is referred to \cite{bote} for the 
complete derivation.

\section{What would have been the 100~m world record in 1988?}
\label{ben979}
On September 24th at the 1988 Seoul Olympics, Ben Johnson of Canada clocked
an astounding 100~metre World Record time of 9.79~s, bettering his previous
record of 9.83~s.  Although this mark (and most of Johnson's other records)
were stricken from the books due to the infamous steroid scandal
which followed, the Seoul mark did not reflect the true potential of this
remarkable athlete.  At about 15~m from the finish line, Johnson looked over
his shoulder to gauge his lead over American Carl Lewis, and then raised his
arm in victory as he coasted through the remained of the race.  The question
lingers: what would have been his World Record time, had he not ``stopped''
at about 85~m?

The model discussed herein can be used to obtain projections of his potential
performances.  Since no instantaneous velocity data is available for this race, 
the parameters in Equation~\ref{mymodel} are selected to match the 10~m splits.
Table~\ref{bensplits} shows these values (obtained from \cite{seoul88}),
along with two sample simulations.
The measured wind-speed of the race was $w = +1.1~$ms$^{-1}$, and the modified
drag area is kept as in Section~\ref{results}.  The elevation of Seoul 
is roughly that of sea level, so altitude corrections will be minimal in this 
case.

Effectively, Johnson could have lowered his previous record by almost
0.1s, a quantum leap in the event.  Depending on the chosen parameters,
the simulations predict times in the range of 9.60-9.62s, which round
to about 9.73-9.75s once reaction time is included. 
In this case, two sets are selected to match the recorded splits, with
the parameter values the same as in Section~\ref{results}, except variation
in the values of $(f_0, f_1)$.  The modified values are
$(f_0 = 6.20; f_1 = 5.16)$ (Simulation~1), and
$(f_0 = 6.18; f_1 = 5.17)$ (Simulation~2).  Note that
the maximum velocity of the simulations do not exceed 12~ms$^{-1}$.
While Johnson may have slightly topped this value in the real race,
he most certainly did not achieve 13.1~ms$^{-1}$, a feat periodically
attributed to him \cite{natpost}.

In 1999, American Maurice Greene reset the World Record to 9.79~s during a 
race in Athens, Greece.  Up to slight variations in altitude, the
city is also at sea level, where the wind gauge read
a calm $+0.1$~ms$^{-1}$.  Adjusting this variable in the
Johnson simulations, the time is scaled down to about 9.665~s
(9.797~s after reaction) for
Simulation~1, and 9.653~s for Simulation~2.
(9.785~s after reaction).  Thus, Johnson's former World Record would
have effectively been equal to the 9.79~s set in 1999 by Greene.

\section{General conclusions and future considerations}
The quasi-physical model considered herein provides a good match to
measured split data, particularly in the drive phase of the race.
The wind and altitude corrections for both men and women
are consistent with those proposed by previous authors, although this
models yields stronger deviations from the norm for variable wind conditions.
If this type of scenario occurs in a real competition, then the times may
appear much faster (or slower) than expect from the measured (time-averaged) wind.

Upon comparison with the previous theoretical estimates and complementary
field studies, it is clear that this model is both a realistic representation
of short-sprint races, as well as a useful tool which can assist in the
physiological and biomechanical study of the sport.

A forthcoming manuscript \cite{me200} will discuss the effects of wind
and altitude assistance in the 200~metre dash.

\pagebreak

\begin{table}[h]
\begin{center}
{\begin{tabular}{c| r r| r r}\hline
d (m) & \multicolumn{2}{c|}{$w = 0.0$~ms$^{-1}$}&\multicolumn{2}{c}{$w = +2.0$~ms$^{-1}$} \\ \hline
10 &    1.708  &      8.800 & 1.705   &     8.840\\
20 &    2.747  &      10.323 & 2.738   &     10.396\\
30 &    3.676  &      11.142 & 3.659   &      11.240\\
40 &    4.554  &      11.584 & 4.529   &     11.710\\
50 &    5.409  &      11.792 & 5.373   &     11.941\\
60 &    6.254  &      11.844 & 6.208   &     12.012\\
70 &    7.100  &      11.787 & 7.041   &      11.970\\
80 &    7.953  &        11.650 & 7.880   &     11.849\\
90 &    8.818  &      11.455 & 8.730   &     11.666\\
100&    9.700  &       11.217 & 9.596  &      11.439\\ \hline
\end{tabular}}
\end{center}
\caption{Model prediction for $H = 0$~m; $(f_0, \sigma) = 
(6.10, 2.22)$; $(f_1, c, \alpha) = (5.15, 0.0385, 0.3225)$; $A_d = 0.002875$.
Maximum velocity 11.845~ms$^{-1}$ at 59.180~m (6.185~s) for $w =0.0$~ms$^{-1}$;
12.012~ms$^{-1}$ at 60.821~m (6.276~s) for $w = +2.0$~ms$^{-1}$.}
\label{modelsplits}
\end{table}

\begin{table}[h]
\begin{center}
{\begin{tabular}{c|c r|c r|c r|c r|c r}\hline
d (m) &\multicolumn{2}{c|}{$H =$~500~m} &\multicolumn{2}{c|}{1000~m} &\multicolumn{2}{c|}{1500~m} & \multicolumn{2}{c|}{2000~m} &\multicolumn{2}{c}{2500~m} \\ \hline
10  &1.708&8.806&1.708&8.811&1.707&8.8148&1.707& 8.820&1.707 & 8.8243 \\
20  &2.746&10.335&2.745&10.345&2.744&10.355&2.743& 10.364&2.742&  10.373 \\
30  &3.674&11.159&3.671&11.174&3.669&11.189&3.667& 11.203&3.665&  11.217 \\
40  &4.550&11.606&4.547&11.627&4.543&11.646&4.540 &11.665&4.537  &11.683 \\
50  &5.403&11.819&5.398&11.845&5.393&11.869&5.388& 11.891&5.383&  11.912 \\
60  &6.246&11.875&6.239&11.905&6.232&11.932&6.226& 11.958& 6.220&   11.980 \\
70  &7.090&11.821&7.080&11.854&7.071&11.885&7.063 &11.914&7.055 & 11.941 \\
80  &7.940&11.687&7.928&11.723&7.917&11.756&7.906 & 11.788&7.896&  11.818 \\
90  &8.803&11.495&8.788&11.533&8.774&11.568&8.761& 11.602&8.749&  11.634 \\
100 &9.681&11.259&9.664&11.298&9.647&11.335&9.631& 11.371&9.617&   11.404 \\ \hline
\end{tabular}}
\end{center}
\caption{Time-velocity (s, ms$^{-1}$) split data for base 9.70~s sprint for
varying altitudes $H$ with $w = 0$~ms$^{-1}$.  Maximum velocities: 
500m: 11.875~ms$^{-1}$ (59.531~m; 6.207~s);  1000m: 11.905~ms$^{-1}$ (59.868~m;
6.228~s); 1500m: 11.932~ms$^{-1}$ (60.176~m; 6.247~s); 2000m: 11.959~ms$^{-1}$
(60.481~m; 6.266~s); 2500m: 11.983~ms$^{-1}$ (60.759~m; 6.283~s).}
\label{altsplits}
\end{table}

\begin{table}[h]
\begin{center}
{\begin{tabular}{c|c r|c r|c r|c r}\hline
d (m) &\multicolumn{2}{c|}{Case~1} &\multicolumn{2}{c|}{Case~2} &\multicolumn{2}{c|}{Case~3} & \multicolumn{2}{c}{Case~4}  \\ \hline
10 &1.708  & 8.800&1.703  &8.865&1.708 &  8.812&1.704 & 8.861 \\
20 & 2.747 &  10.323&2.732& 10.449&2.745& 10.354&2.734 & 10.431 \\
30 & 3.676 &  11.149&3.648& 11.321&3.67 & 11.197&3.652 & 11.282 \\
40 & 4.554 &  11.584&4.511& 11.810&4.543& 11.668&4.519 & 11.747 \\
50 & 5.408 &  11.843&5.348& 12.015&5.39 & 11.908&5.361 & 11.969 \\
60 & 6.247 &  11.979& 6.18& 12.016&6.226& 11.994&6.193 & 12.027 \\
70 &  7.08 &  11.987&7.015& 11.923&7.06 & 11.971&7.026 & 11.967 \\
80 & 7.917 &  11.901&7.859& 11.761&7.899& 11.870&7.867 & 11.822 \\
90 & 8.763 &  11.746&8.717& 11.549&8.746& 11.711&8.72  &11.613 \\
100&  9.621&   11.540&9.592& 11.298&9.608& 11.508&9.59 & 11.355 \\ \hline
$\Delta$ (s) &\multicolumn{2}{c|}{+0.079~s} & \multicolumn{2}{c|}{+0.108~s} 
&\multicolumn{2}{c|}{+0.092~s} & \multicolumn{2}{c}{+0.110~s} \\ 
\end{tabular}}
\end{center}
\caption{Time-averaged +2~ms$^{-1}$ wind-speed corrections. Case columns are
$(t, v)$ split-pairs. $\Delta = 9.700 - t$. }
\label{varwinds}
\end{table}

\begin{table}[h]
\begin{center}
{\begin{tabular}{c|c c}\hline
d (m)  &\multicolumn{2}{c|}{Slow Start} \\ \hline
10 & 1.765  & 8.896 \\
20 &2.788  & 10.486 \\
30 &3.705  & 11.269 \\
40 & 4.575 &  11.677 \\
50 &5.423  & 11.861 \\
60 &6.264  & 11.896 \\
70 &7.107  & 11.826 \\
80 &7.957  & 11.681 \\
90 &8.821  & 11.480 \\
100 &9.701 &  11.238 \\
\end{tabular}}
\end{center}
\caption{
Example $[t (s), v (ms^{-1})]$ simulation for ``slow start'' sprinter (see Figure~\ref{sf}).
}
\label{bailey}
\end{table}

\begin{table}[h]
\begin{center}
{\begin{tabular}{c|c c|c c}\hline
Wind (ms$^{-1}$) &t (s)&$v_{max}$ (ms$^{-1}$)&$d_{max}$ (m) & $\Delta$ (s)  \\ \hline
$2/5 \; (5-t)$&  9.690&11.867&57.534&+0.010\\
$2/5 \; (t-5)$&  9.716&11.823&61.134&$-0.016$\\
$2 \cos(\pi t/5)$&      9.734&11.718&57.490&$-0.034$\\
$-2 \cos(\pi t/5)$&    9.677&11.953&60.293&+0.023       \\
$2 \sin(\pi t/5)$&  9.691&11.885&55.283&+0.009\\
$-2 \sin(\pi t/5)$&9.720&11.821&64.336&$-0.020$\\
$4/5 \; (5-t)$&  9.687&11.889&56.153&+0.013\\
$4/5 \; (t-5)$&  9.739&11.803&63.397&$-0.039$\\
$4 \cos(\pi t/5)$&      9.778&11.575&54.359&$-0.078$\\
$-4 \cos(\pi t/5)$&    9.665&12.043&61.125&+0.035\\
$4 \sin(\pi t/5)$&9.693&11.932&52.749&+0.007\\
$-4 \sin(\pi t/5)$&9.752&11.814&69.150&$-0.052$\\ \hline
\end{tabular}}
\end{center}
\caption{Time-averaged 0~ms$^{-1}$ wind-speed adjustments to base 9.700~s clocking.}
\label{wind0avg}
\end{table}

\begin{table}[h]
\begin{center}
{\begin{tabular}{c|c |c r}\hline
w (ms$^{-1}$) &$ H = 0$~m & 1000~m & 2000~m \\ \hline 
-5 &    9.736&  9.730&  9.724\\
-4 &    9.729&  9.724&  9.717\\
-3 &    9.723&  9.716&  9.710\\
-2 &    9.715&  9.709&  9.703\\
-1 &    9.708&  9.702&  9.696\\
+0 &    ---  &  9.695&  9.689\\
+1 &    9.693&  9.687&  9.683\\
+2 &    9.686&  9.681&  9.676\\
+3 &    9.680&  9.676&  9.672\\
+4 &    9.674&  9.671&  9.667\\
+5 &    9.669&  9.666&  9.663 \\ \hline 
\end{tabular}}
\end{center}
\caption{
Wind and altitude corrected times using ``back-of-the-envelope'' correction
in Equation~(\ref{wind2}).  Base model is 9.700~seconds (0-wind, 0-altitude).
}
\label{boteTable}
\end{table}

\begin{table}[h]
\begin{center}
{\begin{tabular}{c|c |c r|c r}\hline
d (m) &Official Split &\multicolumn{2}{c|}{Simulation~1} &\multicolumn{2}{c}{Simulation~2}\\ \hline
10 &1.70  & 1.698&8.859&  1.698 & 8.864 \\
20 & 2.74 & 2.730&10.396& 2.729& 10.406 \\
30 & 3.63 & 3.652&11.229& 3.650 & 11.242 \\
40 & 4.53 & 4.523&11.685& 4.520& 11.701 \\
50 & 5.37 & 5.370&11.906& 5.365 & 11.924 \\
60 & 6.20 & 6.207&11.969& 6.201& 11.989 \\
70 &  7.04 &7.043&11.922& 7.036 & 11.943 \\
80 & 7.89  &7.886&11.794&7.878& 11.816 \\
90 & 8.76  &8.740&11.607&8.730& 11.629 \\
100&  9.66 &9.611&11.375&9.599& 11.398 \\
$+R$ &9.79~s & \multicolumn{2}{c|}{9.743~s}
&\multicolumn{2}{c}{9.731~s}  \\
\end{tabular}}
\end{center}
\caption{Comparison of theoretical and actual race splits
for Ben Johnson, 1988 Olympic Final. $R = 0.132$~s is the reaction time.}
\label{bensplits}
\end{table}

\pagebreak

\end{document}